\documentclass[preprint,prb,aps,amsmath,amssymb]{revtex4}
\usepackage{graphicx}

\usepackage[english]{babel}
\usepackage{graphicx}
\usepackage{bm}

\begin{document}

\noindent{\large\bf Stimulation of a Singlet Superconductivity in
SFS Weak Links by Spin--Exchange Scattering of Cooper Pairs}

\bigskip\bigskip

\noindent A.~V.~Samokhvalov$^{1,4,*}$, R.~I.~Shekhter$^{2}$,
A.~I.~Buzdin$^{3}$

\bigskip\bigskip

\noindent $^{1}$Institute for Physics of Microstructures, Russian
Academy of Sciences, \\ 603950 Nizhny Novgorod, GSP-105, Russia \\
\medskip
$^{2}$Department of Physics, University of Gothenburg, SE-412 96
G\"oteborg, Sweden\\
\medskip
$^{3}$Institut Universitaire de France and University Bordeaux,
LOMA UMR-CNRS 5798, \\F-33405 Talence Cedex, France \\
\medskip
$^{4}$Lobachevsky State University of Nizhny Novgorod, Nizhny
Novgorod 603950, Russia\\
\medskip
$^{*}$samokh@ipm.sci-nnov.ru
\bigskip\bigskip

\noindent {\bf Josephson junctions with a ferromagnetic metal weak
link reveal a very strong decrease of the critical current
compared to a normal metal weak link. We demonstrate that in the
ballistic regime the presence of a small region with a
non-collinear magnetization near the center of a ferromagnetic
weak link restores the critical current inherent to the normal
metal. The above effect can be stimulated by additional electrical
bias of the magnetic gate which induces a local electron depletion
of ferromagnetic barrier. The underlying physics of the effect is
the interference phenomena due to the magnetic scattering of the
Cooper pair, which reverses its total momentum in the ferromagnet
and thus compensates the phase gain before and after the
spin--reversed scattering.  In contrast with the widely discussed
triplet long ranged proximity effect we elucidate a new singlet
long ranged proximity effect. This phenomenon opens a way to
easily control the properties of SFS junctions and inversely to
manipulate the magnetic moment via the Josephson current.}


\newpage

\bigskip\noindent Mesoscopic properties of nanometer sized conductors are
strongly affected by the injection of correlated electrons (Cooper
pairs) from superconducting electrodes. While propagating through
normal metal such pairs of electrons are able to preserve their
superconducting correlation on mesoscopic lengths providing the
superconducting current flow through SNS (superconductor-normal
metal-superconductor) weak
links~\cite{Barone-Paterno-Physics,Golubov-Kupriyanov-Ilichev-RMP04}.

Time reversal of electronic states forming a Cooper pair is an
important component of the above correlation. Absence of
superconducting pairing interaction in a N part of the
superconducting SNS device opens a possibility of easy external
manipulation of the spin structure of the propagating Cooper
pairs. This offers the means for spintronic manipulation of
superconducting weak links. The most efficient "tailoring" of
Cooper pairs can be achieved by external in-homogeneity located on
a submicron length scale, which is set by superconducting
coherence length, determining the scale of spatial correlation of
paired electrons. Here we demonstrate that a tip of the magnetic
exchange force microscope
(MExFM)~\cite{Kaiser-Wiesendanger-Nature07,Wiesendanger-RMP09},
which induces localized in space magnetic exchange fields can play
the role of such a local probe for a spin state of superconducting
Cooper pairs. Existing experimental evidences of the externally
induced exchange fields $h$ in metals of the order of
few~\cite{Xiong-PRL11,Hamaya-APL07} or even few
tens~\cite{Pasupathy-Sci04} millielectronvolts place the
electronic coupling to such field in a range of energies
comparable with (or even exceeding) superconducting energy scale
$\sim1\, \mathrm{meV}$. We will show that the effect of exchange
induced gating of Cooper pairs leads to a new phenomenon -
stimulation of long- range singlet superconductivity in SFS
(superconductor-ferromagnet-superconductor) weak links.

Magnetic exchange interaction in ferromagnetic metals lifts a
degeneracy with respect to spin orientation of the electrons,
forming a Cooper pair. This leads to different de-Broglie
wavelengths of electrons at Fermi surface for spin-up and
spin-down orientation and produces a modulation of the Cooper pair
wavefunction while propagating along the
ferromagnet~\cite{Demler-PRB97}. As a result an oscillatory
damping of the superconducting ordering is known to appear when a
ferromagnetic ordering occurs in a normal metal link connecting
two S electrodes~\cite{Buzdin-rew}. This phenomenon provides the
basis of the $\pi$-junction realization
\cite{BuzBulPan-JETPL82,Ryazanov-PRL01,Oboznov-Ryazanov-Buzdin-PRL06}.
Considering the quantum mechanics of quasiparticle excitations the
exchange field leads to phase difference $\gamma\sim L/\xi_{h}$
gained between the electron- and hole- like parts of
the total wave function along a path of the length $L$%
~\cite{Blanter-PRB04,Buzdin-PRB11}, where $\xi_{h} = \hbar V_{F} /
2 h$ is a characteristic length determined by the exchange field
($V_{F}$ is the Fermi velocity).

Measurable quantities should be calculated as superpositions of
fast oscillating contributions $e^{i\gamma}$ from different
trajectories and, thus, rapidly vanish with the increasing
distance from the SF boundary. It should be noted though, that a
simple domain structure consisting of two F layers with opposite
orientations of exchange field cancels the phase gain $\gamma$ and
suppresses the destructive effect of an exchange
field~\cite{Blanter-PRB04,Melnikov-Samokhvalov-prl12}. It is clear
therefore that by affecting the spin structure of Cooper pairs,
one influences the strength and spatial distribution of the
proximity effect induced by Cooper pair penetration inside a
ferromagnetic metal. It was suggested before that the Cooper pairs
of electrons with aligned spins (with equal-spin pairs) can be
formed by spatial non-homogeneous
magnetization~\cite{Bergeret-PRL01,Kadigrobov-Shekhter-EL01}.
Since they bind electrons with exactly the same de-Broglie wave
length, these triplet Cooper pairs do not dephase, thereby leading
to long-range proximity effect. Singlet Cooper pairs still become
"filtered out" from spatial transfer of superconducting phase
coherence due to a strong dephasing effect, occurring in long (as
compared with magnetic coherence length $\xi_{h}$) SFS weak links.
To observe such a long ranged triplet superconducting current, the
SFS junctions with a composite F layers comprising three
non-collinear domains, were suggested
theoretically~\cite{Houzet-Buzdin-PRB07,Alidoust-Linder-PRB10,Volkov-Efetov-PRB10}
and realized in recent
experiments~\cite{Robinson-Science10,Khaire-Birge-PRL10}. In such
a case the triplet component is generated by a thin ferromagnetic
domain, located between superconducting lead and a thick central
non-collinear domain. The long ranged Josephson current results
from the interference between these triplet components, generated
by the left and right superconducting leads.

\begin{figure}[t]
\begin{center}
\includegraphics[width=15.0cm]{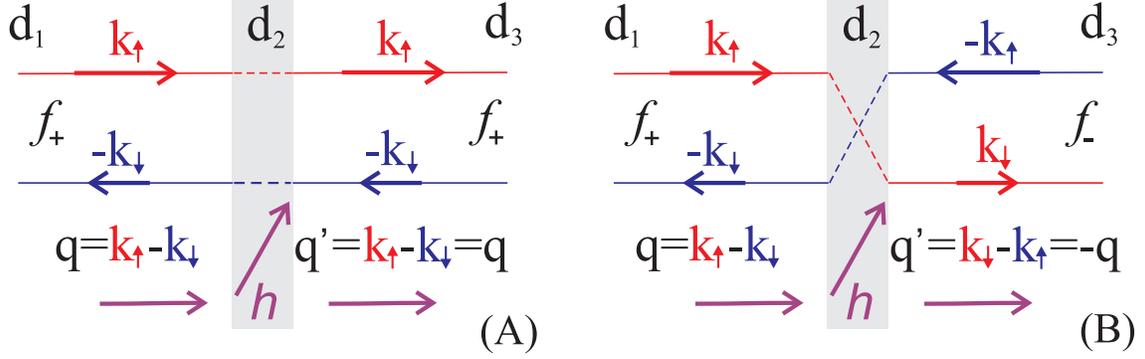}\hspace{2cm}
\end{center}
\caption{(A) A schematic picture of singlet Cooper pair scattering
with no spin--flop transition of electrons. The spin arrangement
of a pair $f_{+}$ ($\mathbf{q} = \mathbf{k}_{\uparrow} -
\mathbf{k}_{\downarrow}$) does not change with respect to the
exchange field: $\mathbf{q}^{\prime}=\mathbf{q}$. (B) A schematic
picture of "exchange" singlet Cooper pair scattering with
spin--flop transition of electrons: $f_{+}\rightarrow f_{-}$. The
spin arrangement of a pair $f_{-}$ was changed with respect to the
exchange field: $\mathbf{q}^{\prime
}=\mathbf{k}_{\downarrow}-\mathbf{k}_{\uparrow}=-\mathbf{q}$. The
scattering domain $d_{2}$ is shown by grey color. The symbols
$f_{\pm}$ indicate the Cooper pairs with zero spin projection and
a reversed spin arrangement (see
\textbf{Methods}).}%
\label{fig1}%
\end{figure}

Here we suggest a new way of manipulating the Cooper pairs flow
through a ferromagnet, which consists in using a well controlled
tip/probe along the supercurrent flow. This corresponds to the
case of the composite F layer with a thin domain, located near the
center of the junction. In contrast with the situation analyzed in
Refs.~\cite{Houzet-Buzdin-PRB07,Alidoust-Linder-PRB10,Volkov-Efetov-PRB10},
the triplet superconducting current is absent for this setup but
becomes possible a long--ranged \textit{singlet} proximity effect.
As we will show the field generated by the probe induces a special
scattering of Cooper pairs which corresponds to exchange spins of
two electrons forming a pair. A schematic picture of such
scattering is shown in Fig.~\ref{fig1}. As we have previously
mentioned two electrons forming a singlet Cooper pair have a
non-zero total momentum
$\hbar\mathbf{q}=\hbar\mathbf{k}_{\uparrow} -
\hbar\mathbf{k}_{\downarrow}$ due to the ferromagnetic exchange
splitting of the spin subbands (~The modulus of the Fermi
wave-vector for electrons with a spin polarized along the field is
larger $\left\vert \mathbf{k}_{\downarrow}\right\vert > \left\vert
\mathbf{k}_{\uparrow }\right\vert$ and $\left\vert \mathbf{q}
\right\vert \sim 1/\xi_{h}$~). The electrons in a singlet Cooper
pair reach the scattering center (spin exchanger) and scatter
their spin so that the new total momentum of the Cooper pair
$\hbar \mathbf{q}^{\prime}$ is either unchanged
$\hbar\mathbf{q}^{\prime }=\hbar\mathbf{q}$ (Fig.~\ref{fig1}A) or
reversed $\hbar\mathbf{q}^{\prime}=-\hbar\mathbf{q}$
(Fig.~\ref{fig1}B) (see the discussion in \textbf{Methods}). In
the first case the spin arrangement of a singlet Cooper pair has
not changed with respect to the exchange field and there remains a total phase gain $\gamma\sim(d_{1}%
+d_{3})/\xi_{h}$ which results in a strong suppression of
proximity due to the destructive trajectory interference. In the
second case the scattered Cooper pair has a reversed spin
arrangement and the total phase gain is $\gamma
\sim(d_{1}-d_{3})/\xi_{h}$ . As a result, at a symmetric position
of the scatterer ($d_{1}=d_{3}$) the total phase gain for a
singlet Cooper pair should be cancelled ($\gamma=0$) and the long
range singlet superconducting proximity in SFS link becomes
possible. %

\begin{figure}[t]
\begin{center}
\includegraphics[width=15cm]{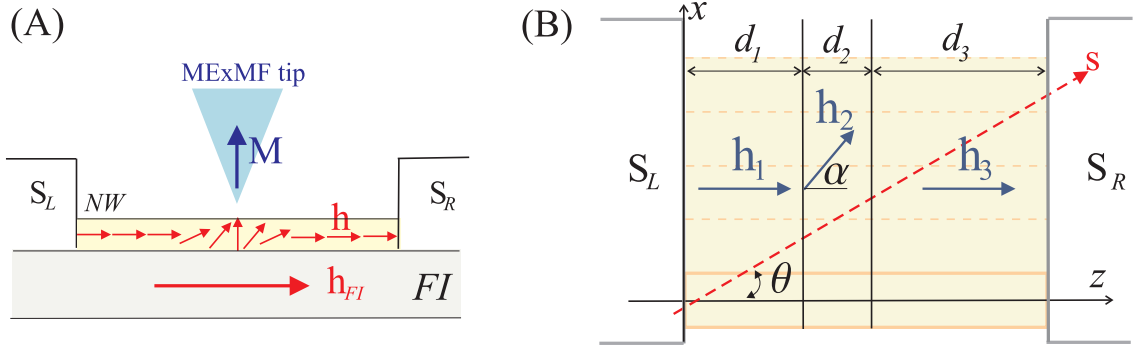}\hspace{2cm}
\end{center}
\caption{(A) The schematic sketch of the SFS constriction under
consideration: normal metal nanowire (NW) in contact with a
superconductor (S) and a ferromagnetic insulator (FI). (B)
Equivalent SFS Josephson junction containing three ferromagnetic
layers (domains) with a stepwise profile of the exchange field
(\ref{eq:1}). Linear quasiparticle trajectory is shown by the red
dashed line.\bigskip}%
\label{fig2}%
\end{figure}

To be more precise, we consider the Josephson transport through a
normal ballistic nanowire (NW) in contact with a ferromagnetic
insulator (FI). The FI turns the NW into an effective ferromagnet
with an exchange field $h$. The schematic picture of the SFS
device is presented in Fig.~\ref{fig2}A. The total length of the
constriction $d = d_{1} + d_{2} + d_{3}$ is assumed to be large
compared to the magnetic coherence length $\xi_{h}=\hbar V_{F} / 2
h$: $d \gg\xi_{h}$. For simplicity we restrict ourselves to the
case of a short junction with $d \ll\xi_{n}$, where $\xi_{n} =
\hbar V_{F} / T_{c}$ is the coherence length of normal metal
($T_{c}$ is the critical temperature of the S layer). The magnetic
tip is assumed to bring on localized in space magnetic exchange
field inhomogeneity which we model by a stepwise profile:
\begin{equation}\label{eq:1}
    h(z) = \left\{\begin{array}[c]{ccc}%
        h \mathbf{z}_{0}, & \mathrm{in\; domains} & d_{1}, d_{3}\\
        h \left( \mathbf{z}_{0} \cos\alpha+ \mathbf{x}_{0}
        \sin\alpha\right) , & \mathrm{in\; domain\;} & d_{2}\,,
    \end{array}\right.
\end{equation}
where $\alpha$ is the angle of the exchange field rotation in the
central domain $d_{2}$ (see Fig.~\ref{fig2}B).
\bigskip

\section*{Results}

The current--phase relation of SFS Josephson junction is
determined by the quasiclassical
relation~\cite{Buzdin-PRB11,Melnikov-Samokhvalov-prl12}
\begin{equation}\label{CPR}
    I=\sum\limits_{n} I_{n} =\sum\limits_{n} a_{n} %
    \sin n\varphi \frac{\langle(\mathbf{n},\mathbf{n}_{F}) \cos n\gamma\rangle}%
    {\langle (\mathbf{n},\mathbf{n}_{F}) \rangle} \ ,
\end{equation}
where $\mathbf{n}$ is the unit vector normal to the junction
plane, $\mathbf{n}_{F}$ is the unit vector along the trajectory,
and $a_{n}$ are the coefficients of the Fourier expansion for the
current--phase relation for superconductor--normal metal junction
of the same geometry. The angular brackets denote the averaging
over different quasiclassical trajectories characterized by a
given angle $\theta$ and a certain starting point at the
superconductor surface, and for 3D constriction looks as
\begin{equation}\label{CPR-brackets}
    \frac{\langle(\mathbf{n},\mathbf{n}_{F}) \cos(n\gamma) \rangle}
    {\langle (\mathbf{n},\mathbf{n}_{F})\rangle} = 2
    \int\limits_{0}^{\pi/2} d\theta \sin\theta\cos\theta\,\cos(n\gamma)\,,
\end{equation}
where $\cos\theta= (\mathbf{n},\mathbf{n}_{F})$. At temperatures
$T$ close to $T_{c}$ the current--phase relation (\ref{CPR}) is
sinusoidal, and the coefficient $a_{1}$ is determined by the
following simple relations~\cite{Melnikov-Samokhvalov-prl12}:
\begin{equation}\label{CPR-a}
    a_{1} = \frac{e T_{c}}{8\, \hbar} N \left(
        \frac{\Delta}{T_{c}}\right) ^{2}\,.
\end{equation}
Here $\Delta$ is the temperature dependent superconducting gap,
The factor $N $ is determined by the number of transverse modes in
the junction: $N = s_{0}^{-1}\int ds \int d\mathbf{n}_{F}
(\mathbf{n}_{F},\mathbf{n}) \sim
S/s_{0}$, where $S$ is the junction cross--section area, and $s_{0}%
^{-1}=(k_{F}/2\pi)^{2}$, where $k_{F}$ is the Fermi momentum.

The phase gain $\gamma$ can be conveniently determined from the
Eilenberger---type equations~\cite{Eilen_1968_eqs} if we use a
standard parametrization~\cite{Champel} of the anomalous
quasiclassical Green function $f=f_{s}+\mathbf{f}_{t}\hat\sigma$,
where $f_{s}$ ($\mathbf{f}_{t}$) singlet (triplet) parts of the
function, respectively, and $\hat\sigma$ is a Pauli matrix vector
in the spin space. The functions $f_{s}$, $\mathbf{f}_{t}$ satisfy
the linearized Eilenberger equations~\cite{Houzet-Buzdin-PRB07}
written for zero Matsubara frequencies
\begin{equation}\label{eq:2}
-i\hbar V_{F} \partial_{s} f_{s}
+2\mathbf{h}\mathbf{f}_{t}=0 \ , \qquad-i\hbar V_{F} \partial_{s}
\mathbf{f}_{t} +2f_{s}\mathbf{h}=0 \ ,
\end{equation}
with the boundary conditions $f_{s}(s=s_{L})=1$,
$\mathbf{f}_{t}(s=s_{L}) =0$ at the left superconducting electrode
(for simplicity we consider the case of the absence of the
barriers at the interfaces). The phase gain $\gamma$ along the
trajectory in the equivalent SFS junction (Fig.~\ref{fig2}B) is
determined by the singlet part of the anomalous quasiclassical
Green function $f_{s}(s=s_{R}) = \cos\gamma$ taken at the right
superconducting electrode~\cite{Melnikov-Samokhvalov-prl12}.
Solving the equations (\ref{eq:2}) for the stepwise profile of the
exchange field (\ref{eq:1}) we find (see \textbf{Methods}):
\begin{align}\label{eq:3}
\cos\gamma & =\cos\delta_{2}
\cos(\delta_{1}+\delta_{3}) -
\cos\alpha\sin\delta_{2} \sin(\delta_{1}+\delta_{3})\nonumber\\
& - \sin^{2}\alpha\, \sin\delta_{1} \sin\delta_{3}
(1-\cos\delta_{2})\, ,
\end{align}
where $\cos\theta= (\mathbf{n},\mathbf{n}_{F})$ and $\delta_{i} = d_{i}%
/\xi_{h} \cos\theta$ ($i=1,2,3$).
Averaging the expression (\ref{eq:3}) over the trajectory
direction $\theta$ and neglecting the terms proportional to
$\xi_{h}/d \ll1$, which decrease just as for the case of
homogeneous ballistic 3D SFS junction, one arrives at the
following long--range (LR) contribution:
\begin{equation}\label{eq:6}
(\cos\gamma)^{LR} = - \frac{1}{2} \sin^{2}\alpha(1 -
\cos \delta_{2})\,\cos2 \delta_{z}\,,
\end{equation}
where $\delta_{z} = z_{0} /\xi_{h} \cos\theta$ and
$z_{0}=(d_{1}-d_{3})/2$ is the shift of the central domain with
respect to the weak link center. So, the long--range component of
the Josephson current at the first harmonic is determined by the
relation:
\begin{equation}\label{eq:8}
I^{LR} \simeq I_{1}^{LR} = a_{1} T_{1}^{LR}
\sin\varphi\,, \qquad T_{1}^{LR} = 2 \int\limits_{0}^{\pi/2}
d\theta\sin\theta\cos\theta \,(\cos\gamma)^{LR}\,.
\end{equation}
For a very thin central domain $d_2 \ll \xi_h$ in the center of
the NW ($z_0 = 0$) one can easily estimate from (\ref{eq:6},
\ref{eq:8}) the critical current of the SFS junction
\begin{equation}\label{eq:8thin}
\mathrm{max} \{I^{LR}\} \approx \frac{I_0}{2}
    \sin^2\alpha \left(\frac{d_2}{\xi_h}\right)^2
    \ln\frac{\xi_h}{d_2}\,,
\end{equation}
where $I_{0} = (e T_{c} N /8 \hbar) \left(
\Delta/T_{c}\right)^{2}$ -- is the critical current of the SNS
junction for zero exchange field ($\gamma=0$). We see that the
long--ranged critical current reaches the maximum at $\alpha =
\pi/2$ and grows with the increase of $d_2$ up to $d_2 \sim
\xi_h$. Our numerical calculations show that it is maximum for
$d_2 \simeq \mathrm{2.5}\,\xi_h$ and may reach $\sim \mathrm{0.7}
I_{0}$. Certainly, the above long--range effect in the first
harmonic describes the properties of the SFS constriction if
contribution of higher harmonics in the current--phase relation
(\ref{CPR}) is negligible. We present the analysis of the second
harmonic effect in the Supplementary information.

\section*{Discussion}

\begin{figure}[ptb]
\begin{center}
\includegraphics[width=12cm]{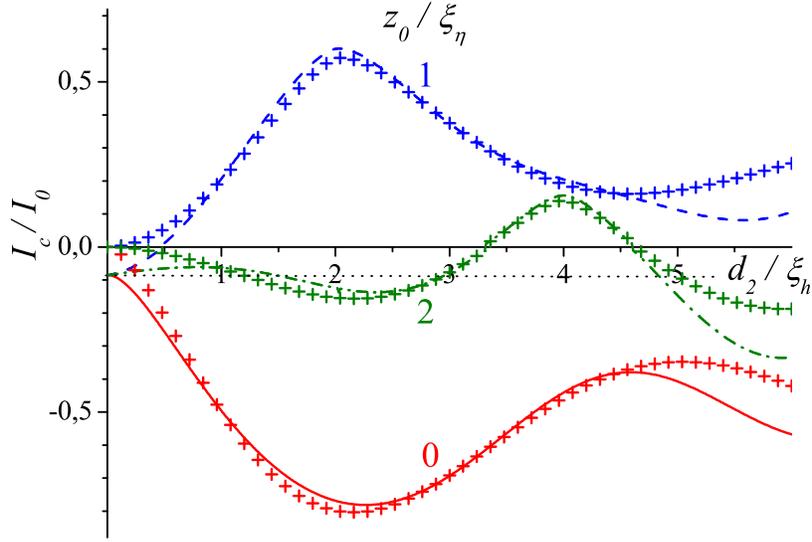}
\end{center}
\caption{The dependence of maximal Josephson current $I_{c}=\mathrm{max}%
\{I_{1}\} $ on the thickness $d_{2}$ of the $90^{o}$ domain
($\alpha=\pi/2$) for different values of the shift of the domain
$z_{0}$: $z_{0} = 0$ - red solid line; $z_{0} = \xi_{h}$ - blue
dashed line; $z_{0} = 2 \xi_{h}$ - green dash-dotted line. Symbols
$+$ show the long-range part of the supercurrent. Dotted line
shows the value of $I_{c}$ in absence of domain $d_{2}$
($\alpha=0$). We have set $T = \mathrm{0.9} T_{c}$; $d = 20
\xi_{h}$ [ $I_{0} = (e T_{c} N /8 \hbar) \left(
\Delta/T_{c}\right)^{2}$ ].}%
\label{fig3}%
\end{figure}

\begin{figure}[ptb]
\begin{center}
\includegraphics[width=12cm]{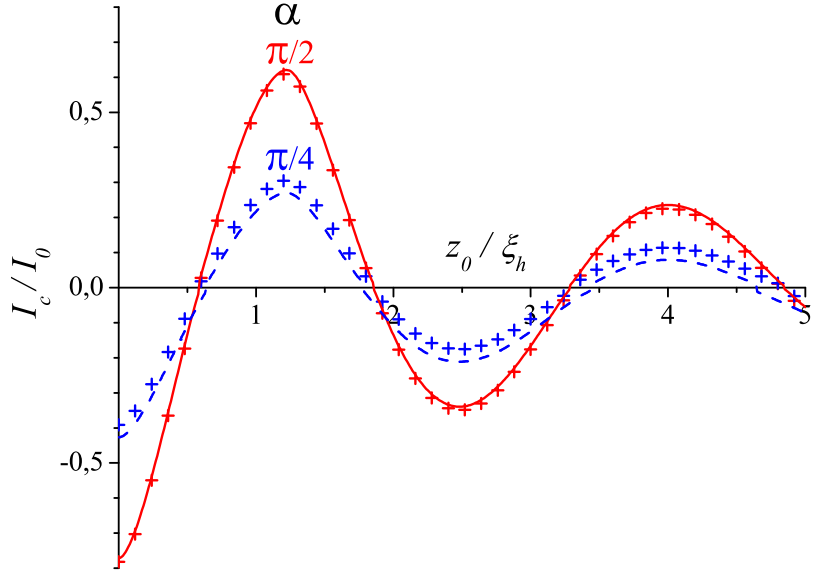}
\end{center}
\caption{The dependence of maximal Josephson current $I_{c}=\mathrm{max}%
\{I_{1}\} $ on the shift of the central domain $z_{0}$ for
different values of the angle $\alpha$: $\alpha=\pi/2$ - red solid
line; $\alpha=\pi/4$ - blue dashed line. Symbols $+$ show the
long-range part of the supercurrent. We have set $T = \mathrm{0.9}
T_{c}$; $d = 20 \xi_{h}$; $d_{2} = 2.5 \xi_{h}$, [
$I_{0} = (e T_{c} N /8 \hbar) \left( \Delta/T_{c}\right) ^{2}$ ].}%
\label{fig4}%
\end{figure}

Figure~\ref{fig3} shows the dependence of the maximal Josephson
current $I_{c}^{LR} = a_{1} T_{1}^{LR}$ on the thickness $d_{2}$
of the $90^{o}$ domain ($\alpha=\pi/2$) for different values of
the shift of the domain $z_{0}$. Naturally, when the thickness of
the central domain $d_{2}$ goes to zero, the long--range effect
disappears. We can see from Fig.~\ref{fig3}, that the long--range
component of the Josephson current $I^{LR}$ coincides
approximately with the total supercurrent across the junction
(\ref{CPR}) until the outer domains are long enough: $d_{1},\,
d_{3} \gg\xi_{h}$. The amplitude of $I^{LR}$ depends
nonmonotonically on the size of the central domain $d_{2}$ and has
the first maximum at $d_{2} \simeq2.5 \xi_{h}$. Interestingly,
that the long--range contribution generates a $\pi-$junction
($I^{LR}$ is negative for zero shift of the domain $z_{0}=0$).
With the shift of the central domain the junction can be switched
from $\pi-$ to $0-$ state.
%
%
Figure~\ref{fig4} shows the dependences of the maximal Josephson
current $I_{c} $ on the position of the central domain $z_{0}$ for
different values of the rotation angle $\alpha$. We may see that
the critical current is very sensitive to the position of the
central domain and the first zero of $I_{1}$ occurs already at
$z_{0} \simeq0.5 \xi_{h}$. %
\bigskip

Here we considered a simple step--like model of magnetization
distribution in F layer. For the very thin central domain $d_{2}
\ll\xi_{h}$, we may easily estimate the long ranged contribution
$(\cos\gamma)^{LR} \sim~ (d_{2} / \xi_{h})^{2}\, (h_{x} / h)^{2}$
which is in accordance with the expression (\ref{eq:6}). For a
general profile of the magnetization it may be convenient to use
the transfer--matrix method (see the Supplementary information).
The smooth (on the scale $\xi_{h}$) profile of the magnetization
decreases the long ranged effect and the proposed mechanism occurs
to be most efficient for $d_{2} \sim\xi_{h}$.

Note that the considered phenomenon should generate the
oscillating potential profile for the magnetic tip $U(z_{0}) \sim-
I_{c}(z_{0}) \cos\varphi$ which depends on superconducting phase
difference across the junction. This opens an interesting
possibility to couple the Josephson current oscillations with
mechanical modes of the tip. On the other hand the same effect can
produce a change of the orientation of the magnetic moment.
Inversely, the precession of the magnetic moment shall modulate
the critical current of the junction and provides a direct
coupling between the superconducting current and the magnetic
moment in the weak link similar to the situation discussed in
Refs.~\cite{buzdin-PRL08,konschelle-buzdin-PRL09}.

The magnetically tunable long-range SFS proximity effect suggested
above has a potential to be an important feature of carbon-based
superconducting weak links. Graphene sheets and carbon nanotubes
are reported to offer a ballistic propagation for electrons on a
micrometer length scale~\cite{Grushina-APL13,Biercuk-CNT}. This
fact together with appearing reports on a gate tunable magnetism
in graphene~\cite{Candini-NLet11,Li-cm13047089} makes all
ingredients of the present theory achievable in experiment.
Another possibility is to use the indium antimonide
($\mathrm{InSb}$) nanowires as a superconducting weak link. The
indium antimonide nanowires, recently used in the experiments to
reveal the signature of Majorana fermions~\cite{Mourik-Sci12},
demonstrated a very high $g-$factor ($g \simeq\mathrm{50}$).
Anomalously large g-factor reported in such wires offers the
possibility to "mimic"\ a ferromagnetic spin-splitting effect of
the order of 10 K by simply applying an external magnetic field of
the order of 0.1 Tesla, and then making such nanowire a suitable
candidate for a weak link to observe the discussed phenomena. Note
that in contrast to the experiments~\cite{Mourik-Sci12} the
magnetic field should be applied along the spin-orbit field axis
to avoid the interference with the spin-orbit effect.

It should be noted, that new additional functionality of the
considered device can be achieved by \textit{electric} biasing of
the magnetic gate~\cite{Haugen-prb08,Semenov-apl07}. In weakly
doped ferromagnetic barriers, such bias ($V_{g}$) alters both the
charge carrier concentration and the Fermi velocity. Choosing a
polarity of electric gating one can create a depletion region
beneath the tip. As a result, both the Fermi velocity $V_{F}$ and
the exchange length $\xi_{h} = \hbar V_{F} / 2 h$ decrease in the
spatial region of the domain $d_{2}$, and the key parameter
responsible for the magnetic exchange scattering $\delta_{2} =
d_{2} / \xi_{h}$ grows. For thin domains ($\delta_{2} \ll1$) the
critical current $I_{c} \sim\delta_{2}^{2}$ increases with the
gate voltage $V_{g}$, and the local depletion of F barrier should
result in the stimulation of the superconductivity. This
nontrivial interplay between electric and magnetic gating effects
can be used to control singlet Josephson current through
ferromagnetic nanowires.

To summarize, we studied the interference phenomena originated by
the spin-exchange scattering in ferromagnetic ballistic weak link
and demonstrated that they provide an efficient way to control the
Josephson current and to couple it with a magnetic moment.


\section*{Methods}


\subsection{Transfer--matrix formalism for Eilenberger Equations}

\begin{figure}[t]
\begin{center}
\includegraphics[width=5.0cm]{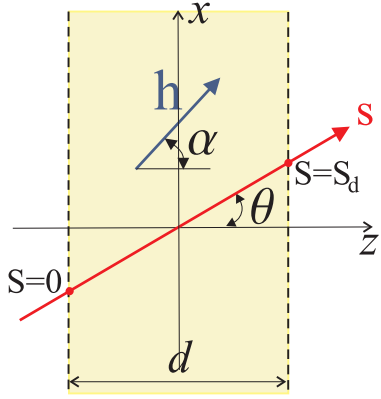}
\end{center}
\caption{Quasiclassical trajectory $\mathbf{s}$ through
homogeneous ferromagnetic layer of a thickness $d$ with an
arbitrary direction of the
exchange field $\mathbf{h}$.}%
\label{fig5}%
\end{figure}

To consider the Josephson transport through ferromagnetic layer
with a non-collinear magnetizations $\mathbf{M}$ and exchange
field $\mathbf{h}$ it is convenient to utilize the
transfer--matrix formalism. For this, we need to solve the
linearized Eilenberger equations written for zero Matsubara
frequencies
\begin{equation}\label{eq:1sm}
-i\hbar V_{F} \partial_{s} f_{s}
+2\mathbf{h}\mathbf{f}_{t}=0 \ , \quad-i\hbar V_{F} \partial_{s}
\mathbf{f}_{t} +2f_{s}\mathbf{h}=0 \ ,
\end{equation}
for the case when the quantization axis is taken arbitrarily in
the ferromagnetic layer of a thickness $d$. We assume that a
quasiclassical trajectory $\mathbf{s}$ and exchange field
$\mathbf{h} = h \left( \mathbf{z}_{0} \sin\alpha+ \mathbf{x}_{0}
\cos\alpha\right) $ lie in the plane ($x,\, z$), as shown in
Fig.~\ref{fig5}.
The trajectory is characterized by a given angle $\theta$ with
respect to the $z$-axis. The triplet part $\mathbf{f}_{t}$ of the
anomalous quasiclassical Green function $f = f_{s} +
\mathbf{f}_{t}\hat\sigma$ consists of two nonzero components and
can be written as $\mathbf{f}_{t} = f_{tx} \mathbf{x}_{0} + f_{tz}
\mathbf{z}_{0}$. Defining the transfer--matrix
$\hat{T}_{\alpha}(d,\theta)$ that relates the components of the
Green function $f(s)$ at the left ($s=0$) and right
($s=s_{d}=d/\cos\theta$) boundaries of the F layer,
\begin{equation}\label{eq:2sm}
\hat{f}(s_{d}) = \left(
    \begin{array}[c]{c}%
        f_{s}(s_{d})\\
        f_{tz}(s_{d})\\
        f_{tx}(s_{d})\\
    \end{array}
    \right)  = \hat{T}_{\alpha}(d,\theta) \left(
    \begin{array}[c]{c}%
        f_{s}(0)\\
        f_{tz}(0)\\
        f_{tx}(0)\\
    \end{array}
    \right)  \,,
\end{equation}
we get the following expression:
\begin{equation}\label{eq:3sm}
\hat{T}_{\alpha}(d,\theta) = \left(
    \begin{array}[c]{ccc}%
        \cos(q s_{d}) & -i \cos\alpha\sin(q s_{d}) & -i \sin\alpha\sin(q s_{d})\\
        -i \cos\alpha\sin(q s_{d}) & \sin^{2}\alpha+ \cos^{2}\alpha\cos(q
        s_{d}) &
        \sin\alpha\cos\alpha\left( \cos(q s_{d})-1\right) \\
        -i \sin\alpha\sin(q s_{d}) & \sin\alpha\cos\alpha\left( \cos(q
        s_{d})-1\right)
        & \cos^{2}\alpha+ \sin^{2}\alpha\cos(q s_{d})\\
    \end{array}\right) \,,
\end{equation}
where $q \equiv1 / \xi_{h} = 2 h / \hbar V_{F} $.

In order to elucidate the peculiarities of the  Cooper pairs
scattering  with a spin-flop transition of electrons it is
convenient to introduce the new functions $f_{\pm}=f_{s}\pm
f_{tz}$ which describes the pairs with zero spin projection and a
reversed spin arrangement. The transfer--matrix $\hat
{T}_{\alpha}(d,\theta)$ can be drastically simplified if the
direction of the exchange field coincides with a spin quantisation
axis $\ \mathbf{z}$. In this case, $\alpha=0$ and
$f_{\pm}(s_{d})=\mathrm{e}^{\mp iqs_{d}}f_{\pm}(0)$,
$f_{tx}(s_{d})=f_{tx}(0).$ Calculating the superconducting current
at the right electrode S$_{R}$ we readily see that it results from
the interference with the singlet component coming from the left
electrode $f_{s}(s_{d})=$ $\left( f_{+}(s_{d})+f_{-}(s_{d})\right)
/2$ (triplet components are irrelevant because the right electrode
provides only the singlet component). The oscillating factors
$\mathrm{e}^{\mp iqs_{d}}$ in $f_{\pm}(s_{d})$ produce, after the
averaging over the trajectories directions ( angle $\theta$ ), a
strong damping of the critical current compared to the normal
metal (where these factors are absent).

Now we may easily understand the mechanism of the long-ranged
proximity effect. Indeed after coming through the first F layer an
extra phase factor appears in $f_{\pm}$ functions (see Fig. 2):
$f_{\pm }(s_{d_{1}})=\mathrm{e}^{\mp iqs_{d_{1}}}f_{s}(0)$. In the
absence the middle
layer, the $f_{\pm}$ at the right electrode would be $f_{\pm}(s_{d_{1}%
}+s_{d_{3}})=\mathrm{e}^{\mp iq(s_{d_{1}}+s_{d_{3}})}f_{s}(0)$ and
the oscillating factors will strongly damp critical current. The
additional non-collinear middle layer $d_{2}$ will mix up the
components $f_{+}$ and
$f_{-}$ - see the matrix (\ref{eq:3sm}) and, for example, $f_{+}(s_{d_{1}%
}+s_{d_{2}})$ in addition to $\mathrm{e}^{-iqs_{d_{1}}}f_{s}(0)$
component will have a $\mathrm{e}^{+iqs_{d_{1}}}f_{s}(0)$
contribution, i. e. \ $f_{+}( s_{d_{1}} + s_{d_{2}} ) = a\,
\mathrm{e}^{-iqs_{d_{1}}}\, +\, b\,\mathrm{e}^{+iqs_{d_{1}}}$. In
fact, namely this mechanism is schematically presented in the Fig.
1(b). Then the resulting \ $f_{+}$ function at the right electrode
should be $f_{+}(s_{d_{1}}+s_{d_{2}}+s_{d_{3}})=a$ $\mathrm{e}^{-iq(s_{d_{1}%
}+s_{d_{3}})}+b\mathrm{e}^{+iq(s_{d_{1}}-s_{d_{3}})}$ and for $d_{1}%
=d_{3\text{ \ \ }}$\bigskip the oscillating factor at the second
term vanishes. This means the emergence of the long-ranged singlet
proximity effect discussed in the present report. Note that the
additional noncollinear F layer $d_2$ may strongly increase the
critical current, provided that it is placed at the center of the
structure.

\bigskip

The transfer--matrix method is very convenient for the calculation
of the Josephson transport through the SFS junction containing
three ferromagnetic layers with a stepwise profile of exchange
field. For this geometry shown in Fig.~\ref{fig2}B of the Letter,
the anomalous quasiclassical Green function
$\hat{f}(s_{R})=(f_{s}(s_{R}),\,f_{tz}(s_{R}),\,f_{tx}(s_{R}))$ at
the right superconducting electrode ($s=s_{R}=d/\cos\theta$) can
be easily expressed via the boundary conditions
$\hat{f}(0)=(1,\,0,\,0)$ at the left superconducting electrode
($s=0$) as follows:
\begin{equation}\label{eq:4sm}
\hat{f}(s_{R})=\hat{T}_{0}(d_{3},\theta)\,\hat{T}_{\alpha}(d_{2},\theta
)\,\hat{T}_{0}(d_{1},\theta)\hat{f}(0)\,,%
\end{equation}
where $d=d_{1}+d_{2}+d_{3}$ is the total thickness of the
ferromagnetic barrier. As a result, the singlet part
$f_{s}(s_{R})$ responsible for the Josephson current through the
junction, can be written in the form (\ref{eq:3}).
%

\bigskip\noindent\textbf{Acknowledgments}\newline The authors thank A.S.
Mel'nikov for stimulating discussions. This work was supported, in
part, by European IRSES program SIMTECH (contract n.246937), by
French ANR grant "MASH", by the Swedish VR, by the Russian
Foundation for Basic Research (n.13-02-97126), and by the program
"Quantum Mesoscopic and Disordered Structures" of the Russian
Academy of Sciences.

\noindent
\section*{Supplementary Information}

{\noindent\bf Supplementary Note 1: Second harmonic contribution}\\
The long--range behavior can be observed for a second harmonic in
the current--phase relation
\begin{equation}\label{eq:1sm}
I=\sum\limits_n I_n =\sum\limits_n I_{c n} \sin n\varphi
\end{equation}
as well. Calculating the $\cos(2\gamma) = 2\cos^2\gamma-1$ we find
the terms which are responsible for a long--range contribution to
the supercurrent:
\begin{eqnarray}
&& (\cos2\gamma)^{LR} = - \sin^2\alpha
\left(1-\cos\delta_2\right) \times                  \label{eq:2sm} \\
        &&\qquad \left[ 1 - \frac{1}{2} \sin^2\alpha \left( 1 - \cos\delta_2
        \right) \left(1 + \frac{1}{2}\cos(4 \delta_z) \right) \right]\,,
        \nonumber
\end{eqnarray}
Averaging over different quasiclassical trajectories for 3D
junction we find a nonvanishing long--range supercurrent at the
second harmonic:
\begin{equation}\label{eq:3sm}
I_2^{LR} = \vert a_{2} \vert T_2\, \sin 2\varphi\,, \quad T_2
\approx -2 \int\limits_0^{\pi/2} d\theta \sin\theta
\cos\theta\,(\cos2\gamma)^{LR}\,,
\end{equation}
where
$$
    a_2 = -\frac{e T_c}{384\, \hbar } N
    \left(\frac{\Delta}{T_c}\right)^4 \,.
$$
\begin{figure}[htb!]
    \includegraphics[width=9cm]{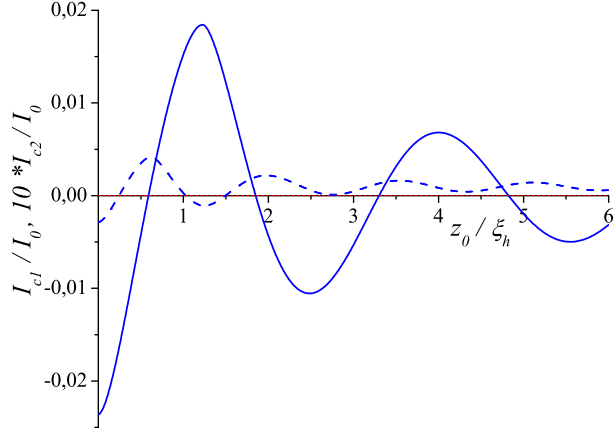}
    \caption{The dependence of the long-range amplitudes of the first $I_{1}$(solid line) and
    the second $I_{2}$ (dashed line) harmonics of the current--phase relation
    (\ref{eq:1sm}) on the shift of the central domain $z_0$.
    We have set $T = \mathrm{0.9} T_c$; $d = 20 \xi_h$; $d_2 = 2.5 \xi_h$; $\alpha=\pi/2$.
    [ $I_0 = (4 e T_c N /\hbar) \left(\Delta/T_c\right)^2$ ].}
    \label{Fig1sm}
\end{figure}
Figure~\ref{Fig1sm} shows the dependences of the function $T_1$
and $T_2$ on position of the central domain $z_0$ for different
values of the rotation angle $\alpha$. We may see that the
critical current is very sensitive to the position of the central
domain and the first zero of $T_1$ occurs already at $z_0 \simeq
0.5 \xi_h$. 

Certainly, the contribution of the second harmonic in the
current--phase relation (\ref{eq:1sm}) is very small, because
usually $|a_2| \ll |a_1|$, except very close to the $0-\pi$
transition ($T_1 = 0$). At this $0-\pi$ transition the
contribution of the second harmonic $I_2$ becomes dominant. For
all considered cases we obtained the positive amplitude of the
second harmonic in the vicinity of these transitions, which means
that they occur discontiguously by a jump between $0-$ and $\pi-$
phase states.

\bigskip\bigskip

{\bigskip\noindent\bf Supplementary Note 2: Arbitrary ferromagnetic barrier}\\
The transfer--matrix formalism can be easily generalized for a
layered ferromagnetic barrier with an arbitrary non-collinear
distribution of the exchange field $\mathbf{h}$ which is described
by the dependence $\alpha(z)$. Splitting the barrier on $N$ thin
layers of the thickness $d_i=z_{i}-z_{i-1}\, (i=1\div N)$ one
consider the exchange field to be constant inside each layer $i$.
So, the transfer--matrix $\hat{T}_{\alpha_i}(d_i,\theta)$ relates
the components of the Green function $f(s)$ at the left
($s_{i-1}=z_{i-1}/\cos\theta$) and right ($s_i=z_i/\cos\theta$)
boundaries of the $i-$layer:
\begin{equation}\label{eq:6sm}
\hat{f}(s_i) = \hat{T}_{\alpha_i}(d_i,\theta)\,
\hat{f}(s_{i-1})\,.
\end{equation}
Application of the transfer--matrix "layer by layer"\ results in
the following relation between the components of the Green
function $f(0)$ and $f(s_R)$ at the left and right superconducting
electrodes, respectively:
\begin{equation}\label{eq:7sm}
\hat{f}(s_R) = \hat{T}_{\alpha_N}(d_N,\theta)\, \ldots\,
\hat{T}_{\alpha_2}(d_2,\theta)\, \hat{T}_{\alpha_1}(d_1,\theta)\,
\hat{f}(0)\,.
\end{equation}

\begin{figure}[hbt!]
    \includegraphics[width=8.0cm]{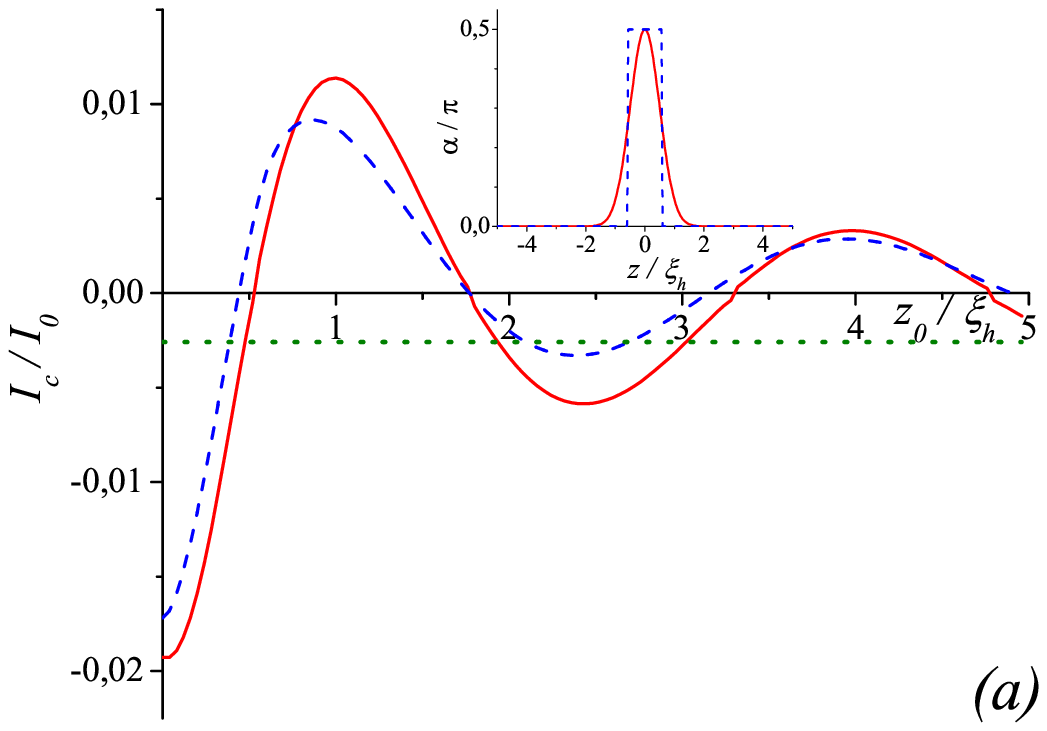}
    \includegraphics[width=8.0cm]{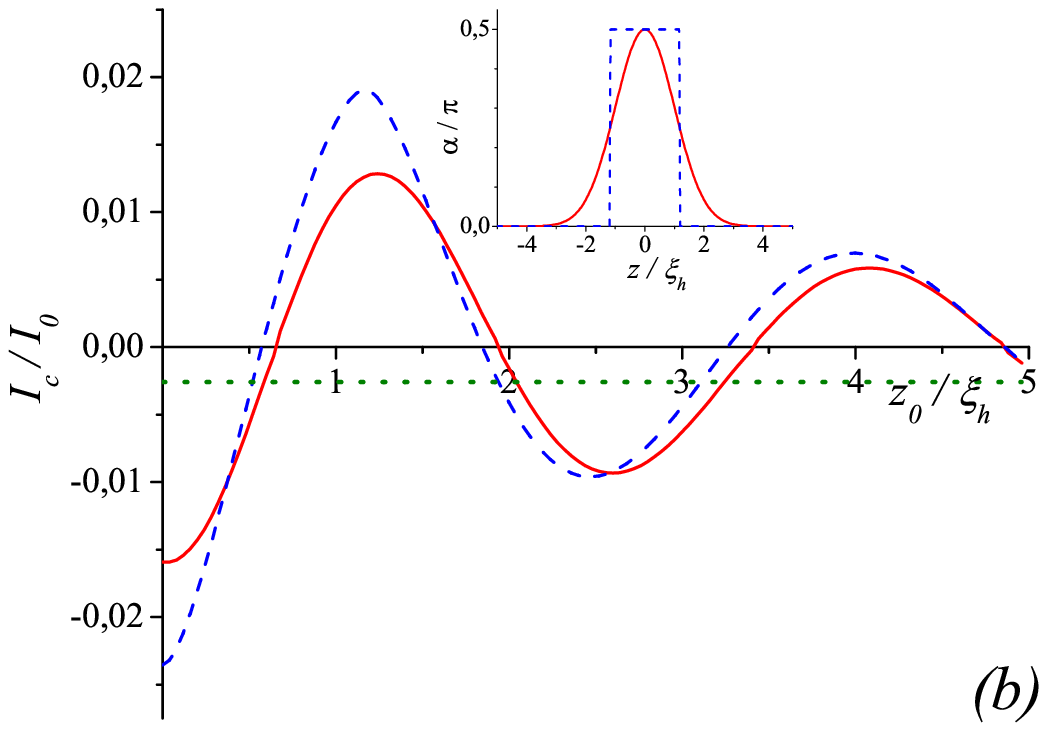}
    \includegraphics[width=8.0cm]{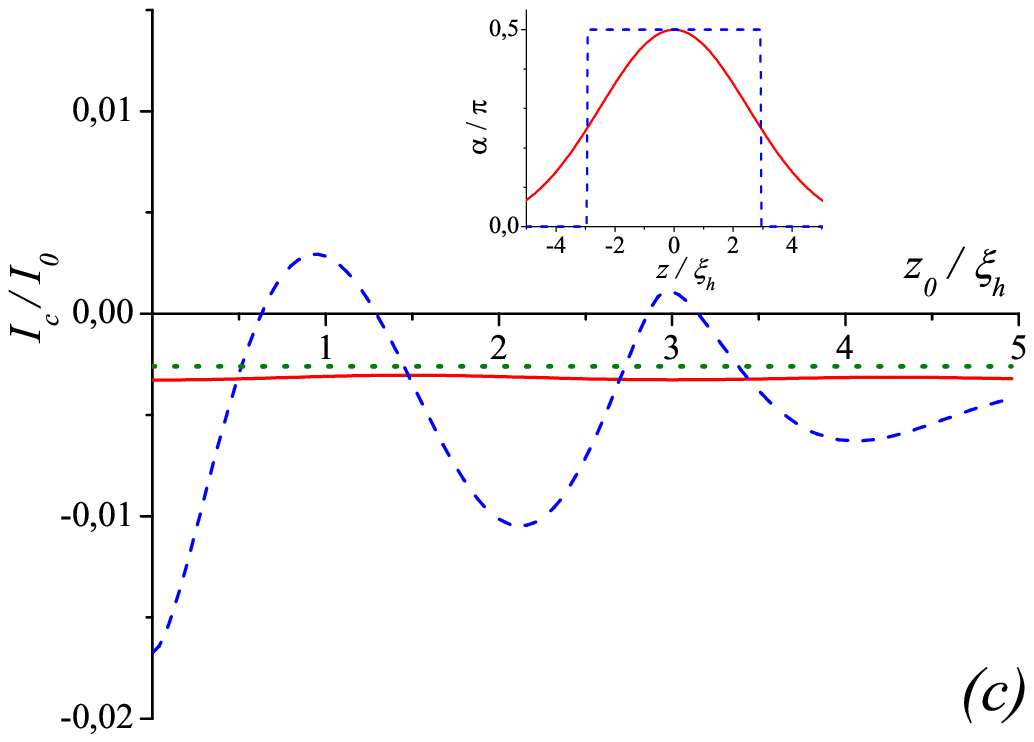}
    \caption{The dependence of the critical current
    $I_c=\mathrm{max}\{I_1+I_2\}$ on
    the shift $z_0$ of the $\mathrm{90^o}-$domain (\ref{eq:8sm})
    ($\alpha=\pi/2$) for different
    values of the width of the transition region $w$ (solid red line):
    (a) $w=1$; (b) $w=2$; (c) $w=5$. The blue dashed line shows
    the dependence of the critical current $I_c$ on
    the shift $z_0$ of stepwise $\mathrm{90^o}-$domain for comparison.
    The dotted line shows the critical current $I_c$
    in the absence of the domain ($\alpha=0$).
    Inset shows the coordinate dependence of the rotation angle $\alpha$
    in the $\mathrm{90^o}-$domain (\ref{eq:8sm})(red solid line)
    and the in the relevant stepwise domain (blue dashed line).
    We have set $d = 20 \xi_h$; $T = \mathrm{0.9} T_c$
    [~$I_0 = (4 e T_c N /\hbar) \left(\Delta/T_c\right)^2$~].}
    \label{fig2sm}
\end{figure}
We apply the described transfer--matrix formalism to study the
effect of smooth in the SFS constriction shown in Fig.2a of the
Letter. As the dependence $\alpha(z)$ we use a draft model of
$\mathrm{90^o}-$domain described by Gaussian funnction:
\begin{equation}\label{eq:8sm}
\alpha(z) =
    \frac{\pi}{2}\exp\left(-\frac{(z-z_0)^2}{2 w^2}\right)\,,
\end{equation}
where $z_0$ is the shift of the domain with respect to the weak
link center, and $w$ describes the width of the domain.
Figure~\ref{fig2sm} shows the dependences of the critical current
of the SFS junction on position $z_0$ of the
$\mathrm{90^o}-$domain (\ref{eq:8sm}) for different values of the
domain width $w$. We may see that the long--range effect seems to
be completely disappeared if $w \gg \xi_h$ (see
Fig.~\ref{fig2sm}c), but it is quite robust for the domain width
smaller than $2-3 \xi_h$  and only weakly depends on the exact
form of the transition region.

\newpage


\bibliographystyle{nature}

\end{document}